\documentclass[conference]{IEEEtran}
\pdfoutput=1

\usepackage{cite}

\usepackage{graphicx}
\usepackage[cmex10]{amsmath}
\usepackage{algorithmic}
\usepackage{array}
\ifCLASSOPTIONcompsoc
  \usepackage[caption=false,font=normalsize,labelfont=sf,textfont=sf]{subfig}
\else
  \usepackage[caption=false,font=footnotesize]{subfig}
\fi
\usepackage{url}

\ifCLASSOPTIONcompsoc 
	\usepackage[caption=false,font=normalsize,labelfont=sf,textfont=sf]{subfig}
\else
	\usepackage[caption=false,font=footnotesize]{subfig}
\fi
\usepackage{todonotes}

\begin{document}
\title{Regression-based Inverter Control for Decentralized Optimal Power Flow and Voltage Regulation}

\author{\IEEEauthorblockN{Oscar Sondermeijer\IEEEauthorrefmark{1}\IEEEauthorrefmark{4},
Roel Dobbe\IEEEauthorrefmark{1}\IEEEauthorrefmark{2}\IEEEauthorrefmark{5}, Daniel Arnold\IEEEauthorrefmark{3},
Claire Tomlin\IEEEauthorrefmark{2} and
Tam\'as Keviczky\IEEEauthorrefmark{4} 
}
\IEEEauthorblockA{\IEEEauthorrefmark{1}These authors contributed equally to this work}
\IEEEauthorblockA{\IEEEauthorrefmark{2}Department of Electrical Engineering \& Computer Sciences, UC Berkeley, Berkeley, USA}
\IEEEauthorblockA{\IEEEauthorrefmark{3}Department of Mechanical Engineering, UC Berkeley, Berkeley, USA}
\IEEEauthorblockA{\IEEEauthorrefmark{4}Delft Center for Systems and Control, Delft University of Technology, Delft, The Netherlands}
\IEEEauthorblockA{\IEEEauthorrefmark{5}Corresponding author: dobbe@berkeley.edu}
}


\maketitle

\begin{abstract}

Electronic power inverters are capable of quickly delivering reactive power to maintain customer voltages within operating tolerances and to reduce system losses in distribution grids. 
This paper proposes a systematic and data-driven approach to determine reactive power inverter output as a function of local measurements in a manner that obtains near optimal results. First, we use a network model and historic load and generation data and do optimal power flow to compute globally optimal reactive power injections for all controllable inverters in the network. Subsequently, we use regression to find a function for each inverter that maps its local historical data to an approximation of its optimal reactive power injection. The resulting functions then serve as \emph{decentralized controllers} in the participating inverters to \emph{predict} the optimal injection based on a new local measurements.
The method achieves near-optimal results when performing voltage- and capacity-constrained loss minimization and voltage flattening, and allows for an efficient volt-VAR optimization (VVO) scheme in which legacy control equipment collaborates with existing inverters to facilitate safe operation of distribution networks with higher levels of distributed generation.

\end{abstract}

\IEEEpeerreviewmaketitle

\section{Introduction} 
%
%
%
%
%

In US distribution grids, voltages are often regulated with an aging infrastructure of capacitor banks, load tap changing transformers and voltage regulators \cite{Meier2006}.
The rapid adoption of distributed generation diversifies power flow and increases the time variability of power and voltage, challenging distribution system operators (DSOs) to revisit their conventional paradigm for reliable operation. 
In areas where photovoltaic (PV) systems are adopted rapidly, such as in Hawaii, the variability leads to increased switching and accelerated wear of legacy equipment and degraded power quality \cite{Stewart2013}.

In transmission networks, the minimization of risk and economic cost are typically handled by optimal power flow (OPF) methods. OPF relies on an extensive and robust communication infrastructure and a very good network model to be executed in real time. As most distribution networks consist of far more nodes with a limited communication infrastructure, it is more challenging to use OPF approaches to control voltage regulating equipment. 
Nevertheless, several works have explored optimal power flow approaches to VAR compensation and Volt-VAR-Optimization \cite{Baran1989A, Baran1989B}. 

Due to rapid adoption of residential scale PV systems, many inverters capable of delivering reactive power support are currently installed. 
An interesting proposition is to use the available inverter capacity, not utilized for real power generation, to control voltage through reactive power injection or absorption. One proposed approach uses an OPF method to infer the optimal reactive power output for all inverters based on a global network objective \cite{Farivar2011,Farivar2013}. This method yields globally optimal power injections and incorporates constraints on voltage and reactive power capacity. A critical assumption is the availability of a communication network to collect measurements from throughout the network and send resulting inverter control signals, which is far from practical for most present day distribution networks. Recent approaches address this issue with distributed \cite{zhang_optimal_2014} and decentralized methods \cite{Zhang2013}. Though advancing in the right direction, the former method still requires communication between all neighboring buses, and the latter suffers from convergence issues.

The second category of methods uses purely local control approaches \cite{Rule21,Smith2011,Turitsyn2010}. These typically use heuristics to adjust reactive power output at each inverter based on the local voltage. These methods have shown promise in their ability to reduce voltage variability, but suffer from extensive tuning which is impractical for larger networks with many inverters. In addition, these methods yield suboptimal control signals and cannot guarantee the satisfaction of critical system constraints.

\subsection{Approach and Contributions}
This work proposes a fully decentralized control scheme for regulating voltage and power flow in distribution grids. It uses offline simulation and regression to understand how groups of inverters can best minimize a collective objective. 
The approach consists of four steps. First, for a specific network with inverters, we collect retrieve data points for all loads and generators for $N$ different scenarios, sampled over an extended period of time and collected by advanced metering infrastructure (AMI). Second, for all $N$ scenarios, we simulate the feeder via a convex OPF problem and determine the globally optimal reactive power injection for all inverters. 
Third, for each individual inverter, we use regression to determine a function that relates its local historical measurement to its optimal reactive power injection. 
Last, we test these functions as controllers in a simulation environment to determine the reactive power injection based on a new local measurement.

We show that our method yields close-to-optimal results for all proposed objectives, and is able to incorporate constraints on voltage, equipment specifications and reactive power capacity. 
The implementation of the regression-based inverter controllers does not need any real-time communication, preventing expensive investments in infrastructure. 
Our method has the ability to collaborate with the existing control equipment, such as load tap changers and capacitor banks. 
\section{Optimal Power Flow}\label{sec:OPF}
This section presents the optimal power flow (OPF) problem that inspires the proposed control approach. It follows the approach in \cite{Low2013} in which the overall optimization problem consists of the objective function, physical power flow equations, additional constraints, and a convex relaxation that enables finding the global optimum. 

\subsection{Power Flow Model}
Solving OPF generally requires a model of line impedances and bus connectivity. Here, a radial network with balanced power flow is assumed, and power flow is described with the well studied DistFlow equations \eqref{eq:BFeqs}, first presented in \cite{Baran1989B}. Consider a graph $\mathcal{G} (\mathcal{N},\mathcal{E})$ with $\mathcal{N}$ the set of nodes and $\mathcal{E}$ the set of edges representing the radial network. Furthermore, $\mathcal{C} \subset \mathcal{N}$ is the subset of $|\mathcal{C}| = I$ nodes that are equipped with controllable PV inverters.  
Capitals $P_i$ and $Q_i$ represent power flow on the line from node $i$, whereas lower case $p_i$ and $q_i$ are the real and reactive power demand at node $i$. Demand is defined as consumption minus generation $p_i = p_i^\text{c} - p_i^\text{g}$, and nodes without a PV system simply have $p^\text{g} = q^\text{g} = 0$. Complex line impedance $r_i + jx_i$ has the same indexing as the power flows. The DistFlow equations use the squared voltage magnitude $v$, defined and indexed at nodes, and the squared current magnitude $\ell$ \eqref{eq:CurEq}. These equations are included as constraints in the optimization problem to enforce that the solution adheres to laws of physics. 
\begin{IEEEeqnarray}{Rl}
P_{i+1} =& P_i - r_i \ell_i - p_{i+1}^\text{c} + p_{i+1}^\text{g} \IEEEyesnumber\label{eq:BFeqs}\IEEEyessubnumber*\\
Q_{i+1} =& Q_i - x_i \ell_i - q_{i+1}^\text{c} + q_{i+1}^\text{g}\\
v_{i+1} =& v_{i} - 2 \left( r_i P_i + x_i Q_i \right) + \left( r_i^2 + x_i^2 \right) \ell_i \\
\ell_i =& \frac{P_i^2 + Q_i^2}{v_i} \IEEEyesnumber\label{eq:CurEq}
\end{IEEEeqnarray}


\subsection{Additional Constraints}
%
The reactive power capacity $\bar{q}[n]$ at time $n$ of an inverter is limited by the total apparent power capacity $\bar{s}$ (constant) minus the real power $\bar{p}[n]$ generated by the PV system at time $n$. As such, the demand of reactive power does not interfere with real power generation.  Therefore, an additional constraint on the inverter reactive power is given by \eqref{eq:InvCap}
\begin{equation}
\left| q_i^\text{g}[n] \right|  \leq \bar{q}_i [n] = \sqrt{\bar{s}_i^2 - (p_i^\text{g}[n])^2}\label{eq:InvCap}
\end{equation}
Here, all inverters are assumed to have $5\%$ overcapacity, so $\bar{s} = 1.05 \bar{p}$, where $\bar{p}$ is the maximum generated real power of the PV system. 

The responsibility of American utilities to maintain service voltage within $\pm 5\%$ of 120$V$ as specified by ANSI Standard C84.1 is expressed as a constraint in the optimization problem \eqref{eq:VoltCons}.
\begin{equation}
\underline{v} \leq v_{i} \leq \overline{v} \label{eq:VoltCons} 
\end{equation} 

The Second Order Cone Programming (SOCP) convex relaxation presented in \cite{Farivar2011} relaxes equality constraint \eqref{eq:CurEq} to inequality \eqref{eq:CurInEq}. 
\begin{equation}
\ell_i \geq \frac{P_i^2 + Q_i^2}{v_i} \label{eq:CurInEq}
\end{equation}

\subsection{Optimization Problem}
The goal of this work is to control inverter reactive power such that the global minimum of a system-wide objective function is obtained. 
In the case study of this paper line losses are minimized and variability in voltage magnitude is reduced. These two goals are formulated with the objective function in \eqref{eq:ObjFun}, with $\gamma$ a trade-off parameter, and $v_{\text{ref}}$ the desired voltage in the network. The optimal power flow problem is given by
\begin{IEEEeqnarray}{Rl}
\min_{\mathbf{z}} \hspace{8pt} &\sum_{i \in \mathcal{E}} r_i \ell_i + \gamma \sum_{i \in \mathcal{N}} | v_i - v_{\text{ref}} | \IEEEyesnumber \label{eq:ObjFun} \\
\text{s.t.} \hspace{8pt} & \eqref{eq:BFeqs}, \eqref{eq:InvCap} - \eqref{eq:CurInEq} \nonumber \\
\mathbf{z} =& \left( v_i, P_i, Q_i, \ell_i, q^\text{g}_i \right) ~\forall i \in \mathcal{N}. \nonumber
\end{IEEEeqnarray}

\section{Regression}\label{sec:Reg}
The goal of regression in this setting is to find a model that approximates the optimal inverter outputs $q_i^\text{g}$ based solely on local measurements. The optimal inverter reactive power outputs are computed for $N$ historical load scenarios, chosen to a representative set of power flow scenarios. Regression is performed for each individual inverter indexed with superscript $(i)$ for $i = 1, \hdots, I$. The $N$ optimal results of $I$ inverters representing all scenarios are used to formulate $I$ vectors $\mathbf{y}^{(i)} = \left[ y^{(i)}[1], ... , y^{(i)}[N] \right]^{\top}$, where $y^{(i)}[n]$ is optimal reactive power output of inverter $i$ at time $n$. The $N$ observations of $K$ input variables $\phi^{(i)}_{k}[n]$ for $n = 1, \hdots, N$ and $k=1,\hdots,K$ are entries of $\mathbf{\Phi}^{(i)} \in R^{N \times K}$. The $K$ input variables of the $n^{\text{th}}$ sample are denoted as $\pmb{\phi}^{(i)}[n] \in R^K$.
\begin{equation}
\mathbf{\Phi}^{(i)} = \begin{bmatrix}
\pmb{\phi}^{(i)\top}[1] \\
\vdots \\
\pmb{\phi}^{(i)\top}[N] 
\end{bmatrix} = 
\begin{bmatrix}
\phi^{(i)}_{1}[1] & \hdots & \phi^{(i)}_{K}[1] \\
\vdots & \ddots & \vdots \\
\phi^{(i)}_{1}[N] & \hdots & \phi^{(i)}_{K}[N]
\end{bmatrix} \label{eq:RegX}
\end{equation}
We use three base local variables at each controllable node $i$:  
\begin{itemize}
	\item real power demand, $\phi^{(i)}_{1}[n] = p_{i}^{c}[n] - p_{i}^{g}[n]$ 
	\item reactive power consumption, $\phi^{(i)}_{2}[n] = q_{i}^\text{c}[n]$ 
	\item reactive power capacity, $\phi^{(i)}_{3}[n] = \bar{q}_{i}[n]$
\end{itemize}
The other $K-3$ columns in the input matrix $\mathbf{\Phi}^{(i)}$ are interaction or quadratic transformations of the base variables, e.g., $\phi^{(i)}_{4}[n] = \phi^{(i)}_{1}[n]\phi^{(i)}_{2}[n]$ and $\phi^{(i)}_{5}[n] = \left(\phi^{(i)}_{1}[n]\right)^2$. 
We use a multiple linear model \eqref{eq:MLRmodel} to relate output $\hat{\mathbf{y}}^{(i)}$ to input matrix $\mathbf{\Phi}^{(i)}$. It is linear in each of the variables, even though these variables themselves can be more nonlinear terms of the base variables. A least squares approach finds the coefficients ${\beta^{(i)}} = \left[ \beta^{(i)}_0, ..., \beta^{(i)}_K \right]$ that minimize the residuals sum of squares \eqref{eq:RSS} given $N$ samples. 
\begin{align}
f(\beta^{(i)},\pmb{\phi}^{(i)}[n]) &= \beta^{(i)}_0 + \beta^{(i)}_1 \phi^{(i)}_{1}[n] + \hdots + \beta^{(i)}_K \phi^{(i)}_{K}[n]. \label{eq:MLRmodel} \\
\text{RSS}( \beta^{(i)} ) &= \sum^N_{n=1} \left(y^{(i)}[n] - f \left({\beta^{(i)}}, \pmb{\phi}^{(i)}[n] \right) \right)^2 \label{eq:RSS}
\end{align}
A hybrid forward- and backward-stepwise selection algorithm selects a subset of input variables \cite{Hastie2009}. The algorithm is initialized with a multiple linear model of the three basic variables only. At each iteration the variable that improves the Bayesian Information Criterion (BIC) value \cite{Schwarz1978} the most, and sufficiently, is added to the model. Subsequently, the variable with the lowest, and sufficiently little, contribution is removed. These two steps are iterated until no variables meet the entrance or exit threshold of the algorithm. The goal of the stepwise selection algorithm is to select the subset of predictors that most accurately predicts the reactive power output of an inverter for unseen input variables. However, the globally best subset is not guaranteed to be found.




\section{Simulation}\label{sec:Sim}
This section first presents the simulation setting used for the case study. Second, the regression-based controller is analyzed and compared to the optimal control scheme. It concludes with an analysis of two regression models.

\subsection{Case study}
\begin{figure}[!t]
\centering
\includegraphics[width=.47\textwidth]{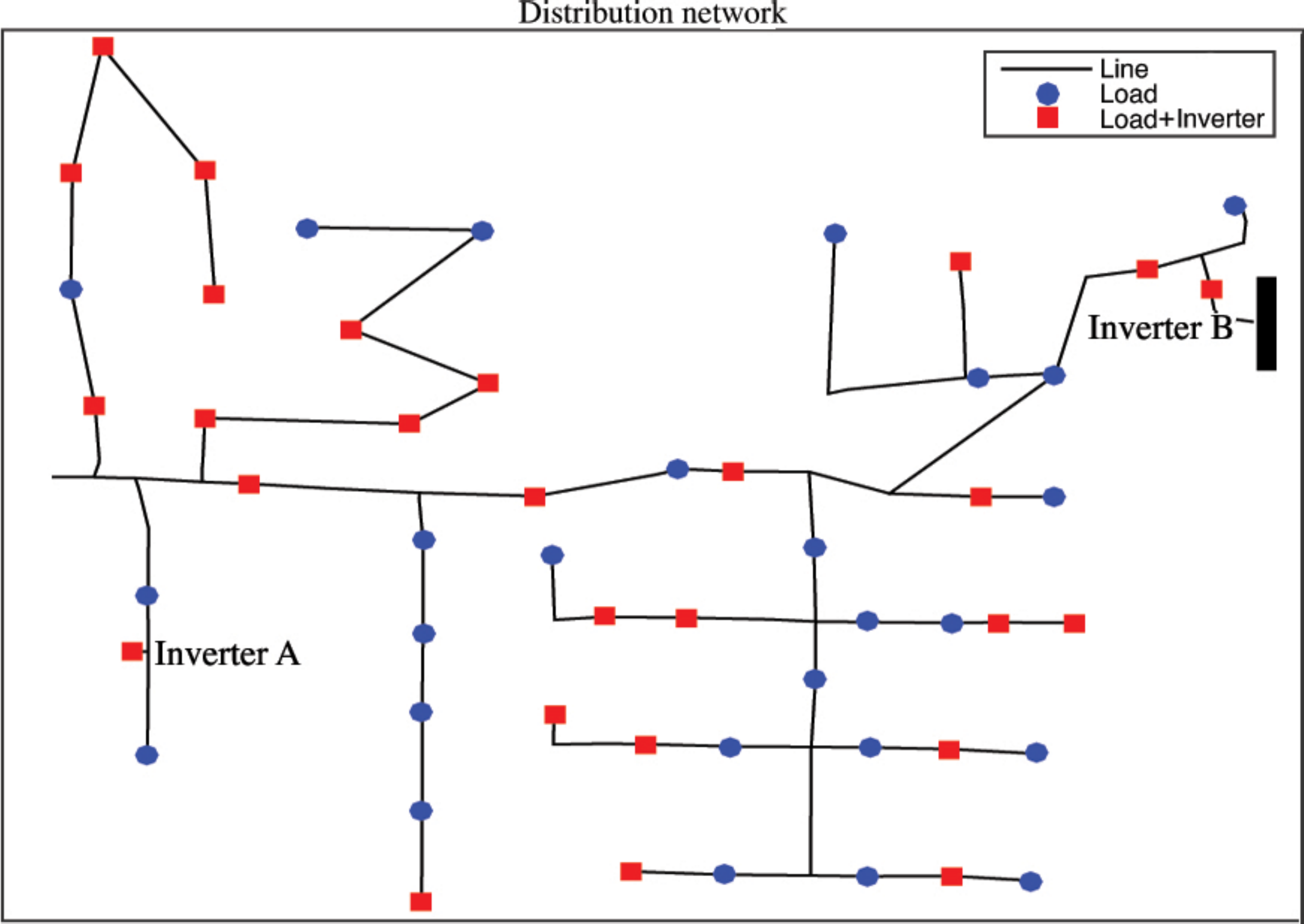}
\caption{One-line diagram of the distribution network used as case study. The substation is located on the far left, locations of loads with PV inverters, and without PV inverters are included, nodes without load nor inverter are omitted for clarity.}
\label{fig:FeederDiag}
\end{figure}

\begin{figure}[!t]
\centering
\includegraphics[width=.5\textwidth]{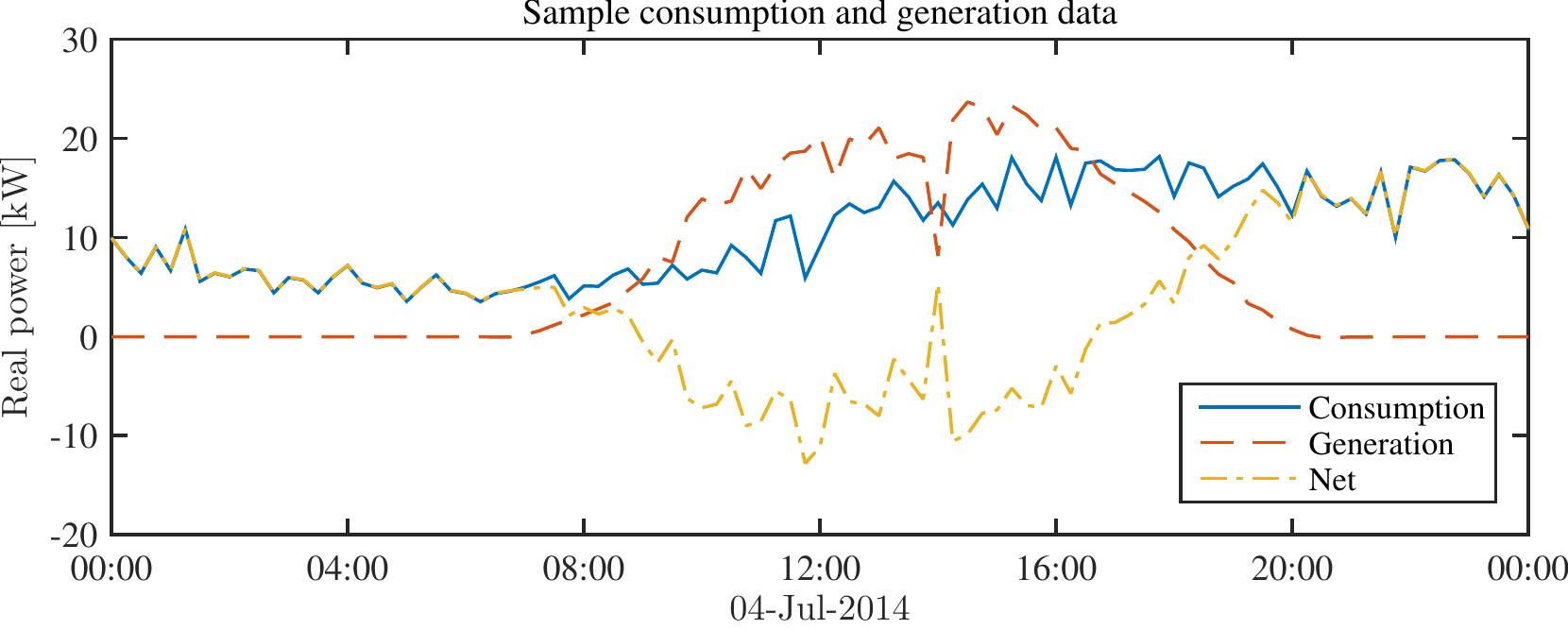}
\caption{Sample real power consumption and PV data at the node of Inverter B on July 4th 2014 from Pecan Street data. The profile is obtained by aggregating data from six individual residences.}
\label{fig:PecanData}
\end{figure}

We evaluate the proposed method on a realistic testbed that is constructed from two independent sources: we construct a 129 node feeder model based on a real distribution feeder from Arizona, Figure \ref{fig:FeederDiag}, and populate this with demand measurements \cite{PecanStreet}. Pecan Street real power consumption and PV generation data with a resolution of 15 minutes is obtained from individual residences in Austin, Texas. For the case study, 50$\%$ of the 53 nodes with loads are randomly selected and equipped with PV installations. Table \ref{tab:FeedProp} includes average values of each line segment's impedance, the average (non-coinciding) peak value of a load, the average power factor of loads, and the average individual power capacity of the 27 inverters.
\begin{table}[!t]
\renewcommand{\arraystretch}{1.3}
\caption{Average Values of Feeder Properties}
\label{tab:FeedProp}
\centering
\begin{tabular}{c c c c}
\hline
Line Impedance & Peak Load & Power Factor & Inverter Capacity \\
\hline\hline
 $0.10 + j 0.07 \Omega$ & 26 kVA & 0.92 & 24kVA \\
\hline
\end{tabular}
\end{table}
Optimization problems are solved with sampled load and PV generation data to retrieve the optimal reactive power output of all inverters. The data is separated into training and validation data: July 5th - 31st 2014 is used as training data for the regression; the obtained models are simulated on validation data from July 4th. The real power consumption and PV generation profile at inverter B is shown in Figure \ref{fig:PecanData} to illustrate typical generation and consumption characteristics on July 4th 2014. In general, Pecan Street data is randomly selected and aggregated to match the typical load for each bus as specified in the feeder model, e.g., the load at Inverter B corresponds to six residences.

The proposed control approach is simulated and compared to two other approaches: a situation where inverter reactive power capacities are not utilized, and the approach described in \cite{Smith2011} where inverters are operated at a constant, non-unity, power factor. In our context, we chose to tune the inverters to operate at lagging (generating) power factor of 0.9 to reduce losses. 
In addition to the comparison, we extend our method to show it is capable of collaboration with a load tap changer (LTC). We design a scheme in which the inverters operate with controllers to flatten the voltage throughout the feeder, and then adjust the turn ratio of the LTC at the substation, to safely lower the voltages throughout the feeder.
Hence, we consider four approaches: 
\begin{description}
\item[a)] no reactive power support
\item[b)] constant power factor inverter operation
\item[c)] regression-based inverter control (reactive power)
\item[d)] collaboration between regression-based inverter control and substation LTC
\end{description}
\subsection{Control simulation results}
%

\begin{figure}[!t]
    \centering
    \includegraphics[width=.5\textwidth]{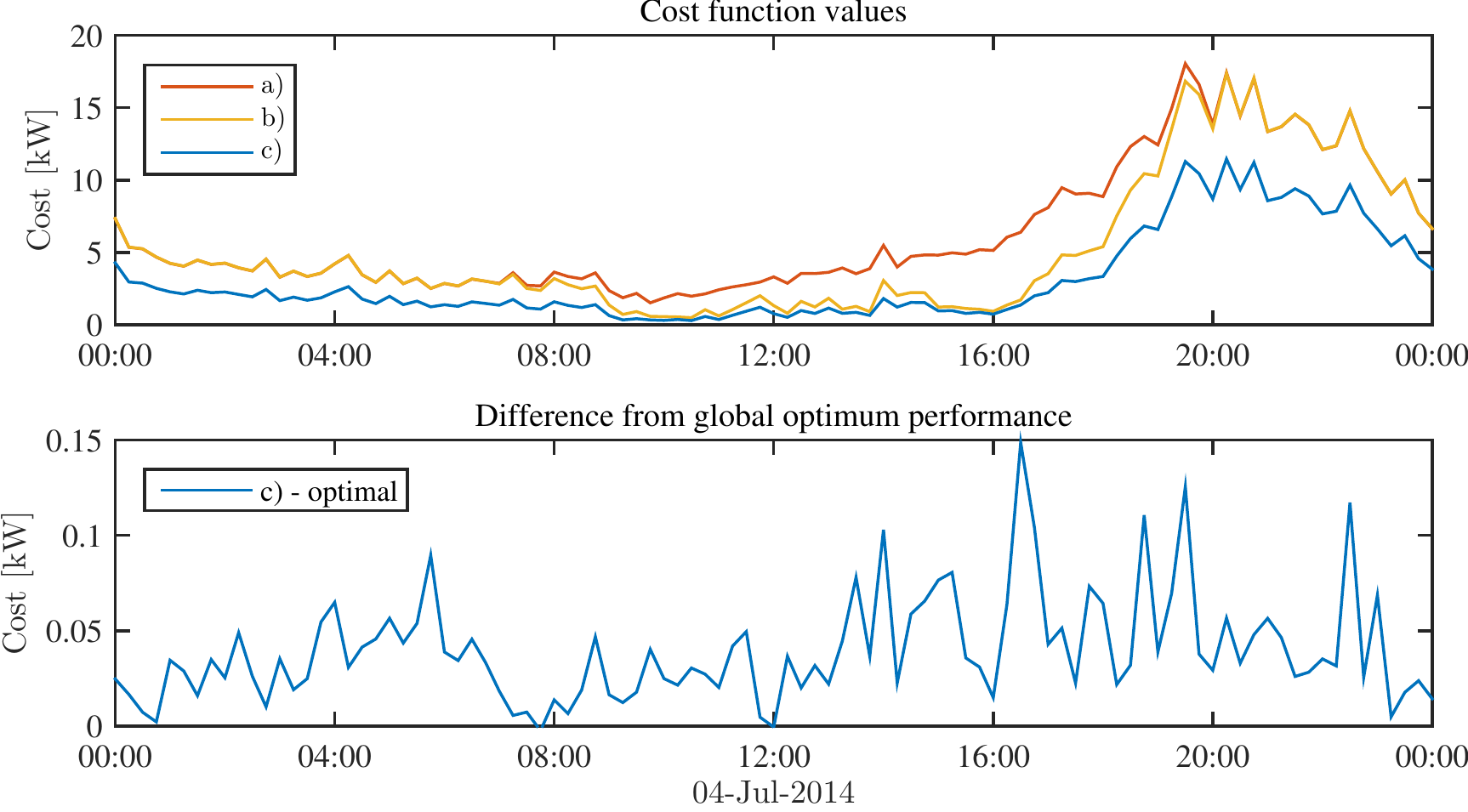}
    \caption{Objective function values of a), b), and c) in upper figure. Difference between regression-based control, c), and optimal objective function values in lower figure, which is two orders of magnitude smaller than the optimal objective function.}
    \label{fig:JValSim}
    \end{figure}

\begin{figure}[!t]
    \centering
    \includegraphics[width=.5\textwidth]{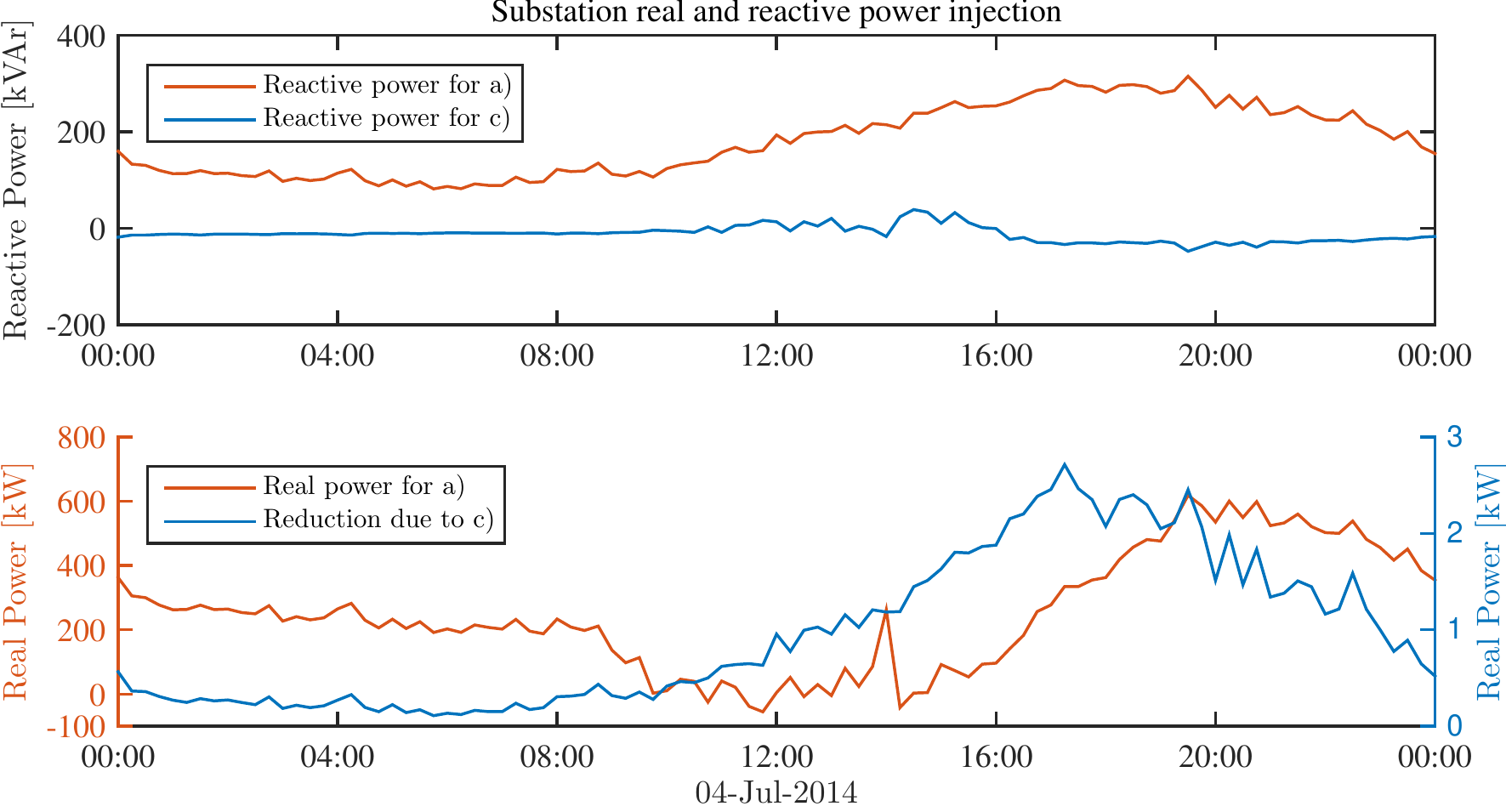}
    \caption{Substation reactive power injection for approach a) (red line), and approach c) (blue line) in upper figure. The reduction in feeder head reactive power injection is 91--393 kVAr. Substation real power injection under approach a) (red line) is shown on the left axis of the lower figure, the right axis shows the reduction in real power injection, which is 0--3 kW.}
    \label{fig:FeederHead}
    \end{figure}

Figure \ref{fig:JValSim} compares the objective function values \eqref{eq:ObjFun} for approaches a), b), and c). Compared to the situation of approach a), both approach b) and c) have beneficial effect on the objective function. However, approach c) achieves the best performance at all times. Approach b) generates reactive power proportional to the real power output of a PV system. This is reflected in a lack of control during hours when PV power is not available.
Between 12.00 and 20.00 the objective function value of approach a) increases rapidly, which is caused by the transition from peak real power generation to peak real consumption, Figure \ref{fig:PecanData} and \ref{fig:FeederHead}. The objective function value of approach c) also increases, but significantly less. 

The lower figure of Figure \ref{fig:JValSim} shows the difference between the global optimal objective function values and the regression-based objective function values. The maximum difference is 1.6$\%$ of the optimal function value, whereas the average difference is $0.15\%$ of the objective function value. Hence, approach c) achieves near-optimal performance.

Figure \ref{fig:FeederHead} compares substation real and reactive power injection in the system under approaches a) and c). Whereas the substation injects an average of 180 kVAr for approach a), the substation absorbs an average of 12 kVAr with the near-optimal reactive power provision of approach c). This implies that all reactive power consumption is locally sourced from distributed PV systems. It improves the power factor at the substation, reduces reactive power loss, and reduces resistive line losses with 0--3 kW. Furthermore, between 10:00--16:00 real power demand is predominantly supplied by PV systems, which results in low objective function values in Figure \ref{fig:JValSim}. 

%

\begin{figure}[!t]
    \centering
    \includegraphics[width=.5\textwidth]{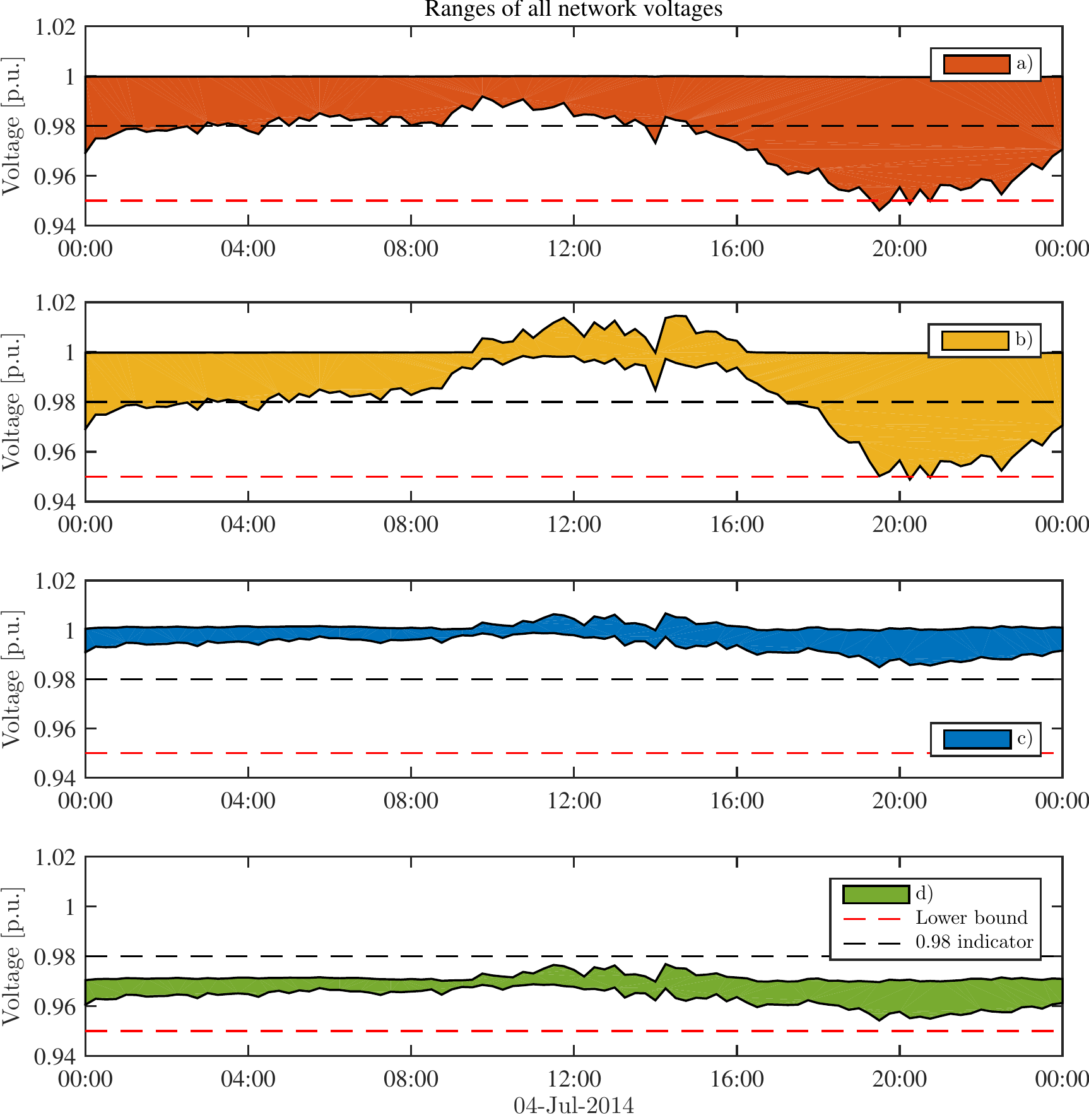}
    \caption{All network voltages for approaches a), b), c), and d). Colored planes represent the range between the maximum and minimum voltages at any node in the network. Lower voltage bound is indicated with red dashed line, an additional indicator is included as a black dashed line at 0.98 p.u., and substation voltage is assumed to be constant at 1 p.u..}
    \label{fig:VoltSim}
    \end{figure}

Figure \ref{fig:VoltSim} presents voltages at all nodes in the network for all  four approaches, indicated with colored surfaces.
For approach a), the voltage drop in the system is smallest between 10:00--16:00, when most real power demand is supplied by PV systems, Figure \ref{fig:FeederHead}. During these hours, the combination of real power injection of PV systems and reactive power generation of approach b) prompts a voltage rise in the distribution feeder. Alternatively, approach c) achieves system voltages that are close to the nominal value of 1 p.u., and simultaneously reduces losses as implied by the lowest objective function value of Figure \ref{fig:JValSim}. The transition from peak PV generation to peak consumption between 12.00 and 20.00 causes the system voltage to change dramatically without regression-based control. For approaches a) and b), the lower bound of the ANSI standard is violated in the evening if traditional voltage regulators are not operated. The effect of approach c) is obvious at these times: reactive power generated by inverters reduce voltage drop in the system and reduces losses, as depicted in Figure \ref{fig:VoltSim} and \ref{fig:JValSim}. Approach d) exploits the reduced voltage variability achieved with approach c), and allows the substation to lower the overall voltage. Finally, the voltages in approach a) and b) show a significant change at 14:00. This is caused by a temporary and sudden reduction of PV generation of approximately 60\% at 14:00, that is best seen in the generation of Figure \ref{fig:PecanData} and the substation real power injection in Figure \ref{fig:FeederHead}. The regression-based controller acts appropriately to this sudden change; no significant change in system voltages is observed in Figure \ref{fig:VoltSim} for approach c). 

\subsection{Regression accuracy}
\label{subsec:regresacc}
%
In our case study, we determined regression models for 27 different inverters. Table \ref{tab:InvMdl} shows regression results for inverters A and B (both indicated in Figure \ref{fig:FeederDiag}). The first two columns present the  features selected by the stepwise regression and the values for the $\beta$-coefficients in \eqref{eq:MLRmodel}. The third column shows the standard error of the estimate, and the fourth lists p-values, which here means the probability that coefficient is zero. A p-value of 0.1 implies a 10\% chance that the corresponding coefficient is zero. Note how the stepwise regression approach results in two clearly different models. The results show that reactive power output of inverter A predominantly depends on $\phi^{(A)}_2$ (reactive power consumption), while the output of inverter B is strongly related to $\phi^{(B)}_3$ (reactive power capacity) and $\phi^{(B)}_2$ is irrelevant. This example illustrates that optimal reactive power output of two inverters can have a completely different structure. Therefore, effective design of controllers based on local measurements is challenging, and can clearly benefit from a data driven approach. 
%
\begin{table}[!t]
\renewcommand{\arraystretch}{1.3}
\caption{Normalized regression coefficients for inverter A and B. (Explanation in Subsection \ref{subsec:regresacc})}
\label{tab:InvMdl}
\centering
\begin{tabular}{c c c c || c c c c}
\hline
A & Est. & SE & p-value &  B & Est. & SE & p-value \\
\hline\hline
offset	&	0.02	&	0.01	&	0.01				&	offset	&	0.02	&	0.00	&	$8.2 e^{-4}$ \\
$\phi_1$	&	0.37	&	0.01	&	$1.8 e^{-164}$		&	$\phi_1$	&	0.04	&	0.00	&	$3.2e^{-22}$ \\
$\phi_2$	&	0.77	&	0.01	&	0				&	$\phi_3$	&	0.96	&	0.01	&	0 \\
$\phi_3$	&	-0.21	&	0.02	&	$8.0e^{-24}$		&	$\phi_1 \phi_3$	&	0.06	&	0.01	&	$1.2e^{-12} $\\
$\phi_1 \phi_2$	&	0.17	&	0.01	&	$1.8e^{-67}$		&	$\phi_1^2$	&	-0.03	&	0.00	&	$3.9e^{-14}$\\
$\phi_1^2$	&	-0.06	&	0.01	&	$1.4e^{-12}$		&	$\phi_3^2$	&	-0.03	&	0.01	&	$6.1e^{-9}$\\
\hline
\end{tabular}
\end{table}



\section{Conclusion}

Regression-based control presents a data-driven solution for the design of a decentralized Volt-VAR optimization scheme. The method mimics an optimal power flow approach and achieves near-optimal performance with inverters providing reactive power compensation based solely on local measurements. Case study results show reduced losses and reduced voltage fluctuation for a feeder with high penetration of PV systems. This prevents excessive switching and accelerated wear of legacy control and protection equipment, a major concern for DSOs. As such, our control strategy allows DSOs to keep operating their legacy equipment in the specified regime. Furthermore, the method enables collaboration between inverters and substation transformer: inverters maintain a flat voltage profile throughout the feeder, while tap changing transformers lower the voltage without violating constraints.

This case study serves as a proof-of-concept and considered a specific objective function, however the extension to include other and more elaborate convex objectives is straightforward. The regression approach can be applied to OPF with different objective functions to more accurately reflect specific goals for other feeders. 
An interesting next step is to analyze the inverter capacity needed to balance the negative effects for anticipated higher levels of DG penetration.

\section*{Acknowledgment}
The authors would like to thank Alexandra von Meier and Duncan Callaway for useful feedback and facilitating discussions with utilities, Kyle Brady and Michael Sankur for helping with the collection and organization of feeder models and load and PV generation data sets, and John Mead and his team at PG\&E Applied Technology Services for fruitful discussions on VVO and testbeds.

\bibliographystyle{IEEEtran}
\bibliography{PaperPes_OSondermeijer_v3}

\begin{thebibliography}{10}
\providecommand{\url}[1]{#1}
\csname url@samestyle\endcsname
\providecommand{\newblock}{\relax}
\providecommand{\bibinfo}[2]{#2}
\providecommand{\BIBentrySTDinterwordspacing}{\spaceskip=0pt\relax}
\providecommand{\BIBentryALTinterwordstretchfactor}{4}
\providecommand{\BIBentryALTinterwordspacing}{\spaceskip=\fontdimen2\font plus
\BIBentryALTinterwordstretchfactor\fontdimen3\font minus
  \fontdimen4\font\relax}
\providecommand{\BIBforeignlanguage}[2]{{%
\expandafter\ifx\csname l@#1\endcsname\relax
\typeout{** WARNING: IEEEtran.bst: No hyphenation pattern has been}%
\typeout{** loaded for the language `#1'. Using the pattern for}%
\typeout{** the default language instead.}%
\else
\language=\csname l@#1\endcsname
\fi
#2}}
\providecommand{\BIBdecl}{\relax}
\BIBdecl

\bibitem{Meier2006}
A.~von Meier, \emph{{Electric Power Systems: A Conceptual Introduction}}.\hskip
  1em plus 0.5em minus 0.4em\relax John Wiley {\&} Sons, 2006.

\bibitem{Stewart2013}
E.~Stewart, J.~Macpherson, S.~Vasilic, S.~Ramon, D.~Nakafuji, and T.~Aukai,
  ``{Analysis of High-Penetration Levels of Photovoltaics into the Distribution
  Grid on Oahu},'' no. May, 2013.

\bibitem{Baran1989A}
M.~E. Baran and F.~F. Wu, ``{Network reconfiguration in distribution systems
  for loss reduction and load balancing},'' \emph{IEEE Transactions on Power
  Delivery}, vol.~4, no.~2, pp. 1401--1407, 1989.

\bibitem{Baran1989B}
------, ``{Optimal Sizing of Capacitors Placed on a Radial Distribution
  System},'' \emph{IEEE Transactions on Power Delivery}, vol.~4, no.~1, pp. 735
  -- 743, 1989.

\bibitem{Farivar2011}
M.~Farivar, C.~R. Clarke, S.~H. Low, and K.~M. Chandy, ``{Inverter VAR control
  for distribution systems with renewables},'' \emph{2011 IEEE International
  Conference on Smart Grid Communications, SmartGridComm 2011}, pp. 457--462,
  2011.

\bibitem{Farivar2013}
M.~Farivar and S.~Low, ``{Equilibrium and dynamics of local voltage control in
  distribution systems},'' in \emph{52nd IEEE Conference on Decision and
  Control}.\hskip 1em plus 0.5em minus 0.4em\relax IEEE, dec 2013, pp.
  4329--4334.

\bibitem{zhang_optimal_2014}
B.~Zhang, A.~Lam, A.~Dominguez-Garcia, and D.~Tse, ``An {Optimal} and
  {Distributed} {Method} for {Voltage} {Regulation} in {Power} {Distribution}
  {Systems},'' \emph{IEEE Transactions on Power Systems}, vol.~PP, no.~99, pp.
  1--13, 2014.

\bibitem{Zhang2013}
B.~Zhang, A.~D. Dominguez-Garcia, and D.~Tse, ``{A local control approach to
  voltage regulation in distribution networks},'' in \emph{2013 North American
  Power Symposium (NAPS)}.\hskip 1em plus 0.5em minus 0.4em\relax IEEE, sep
  2013, pp. 1--6.

\bibitem{Rule21}
\BIBentryALTinterwordspacing
``{Rule 21 Smart Inverter Working Group Technical Reference Materials}.''
  [Online]. Available:
  \url{http://www.energy.ca.gov/electricity_analysis/rule21/}
\BIBentrySTDinterwordspacing

\bibitem{Smith2011}
J.~W. Smith, W.~Sunderman, R.~Dugan, and B.~Seal, ``{Smart inverter volt/var
  control functions for high penetration of PV on distribution systems},''
  \emph{2011 IEEE/PES Power Systems Conference and Exposition}, pp. 1--6, mar
  2011.

\bibitem{Turitsyn2010}
K.~Turitsyn, P.~Sulc, S.~Backhaus, and M.~Chertkov, ``{Distributed control of
  reactive power flow in a radial distribution circuit with high photovoltaic
  penetration},'' in \emph{IEEE PES General Meeting}.\hskip 1em plus 0.5em
  minus 0.4em\relax IEEE, jul 2010, pp. 1--6.

\bibitem{Low2013}
S.~H. Low, ``{Convex Relaxation of Optimal Power Flow: A Tutorial},'' vol.
  2013, 2013, pp. 1--15.

\bibitem{Hastie2009}
T.~Hastie, R.~Tibshirani, and J.~Friedman, \emph{{The Elements of Statistical
  Learning: Data Mining, Inference, and Prediction}}, 2009, vol.~27, no.~2.

\bibitem{Schwarz1978}
G.~Schwarz, ``\BIBforeignlanguage{EN}{{Estimating the Dimension of a Model}},''
  \emph{\BIBforeignlanguage{EN}{The Annals of Statistics}}, vol.~6, no.~2, pp.
  461--464, mar 1978.

\bibitem{PecanStreet}
\BIBentryALTinterwordspacing
``{Pecan Street Inc. Dataport 2014}.'' [Online]. Available:
  \url{https://dataport.pecanstreet.org/}
\BIBentrySTDinterwordspacing

\end{thebibliography}
%
%
%

\end{document}